\documentclass[a4paper,11pt,onecolumn,twoside]{article}
\usepackage{amsmath,amsfonts,amssymb}
\usepackage{graphicx}
\usepackage{mathptmx}
\usepackage{booktabs}
\usepackage[labelfont=bf]{caption}
\usepackage{indentfirst}
\usepackage{caption}
\usepackage{enumitem}
\usepackage{subfigure}
\usepackage{float}
\usepackage{emptypage} 
\usepackage{fancyhdr}
\usepackage[marginal]{footmisc}
\usepackage{setspace}
\usepackage{listings}
\usepackage{xcolor}
\usepackage{pifont}

\newcommand{\upcite}[1]{
  \textsuperscript{\cite{#1}}
}

\title{\textbf{An Hydrodynamic-Instabililty Based Approach towards the Oscillation and Coupling of Candle Flames}}
\author{
Xiaoyue Ma\footnote{201830014@smail.nju.edu.cn}
\\[2pt]
{\small \textit{(2020, School of Physics, Nanjing University, Nanjing 210046)}}\\[6pt]
Tianyi Gu
\\[2pt]
{\small \textit{(2020, Kuang Yaming Honors School, Nanjing University, Nanjing 210046)}}\\[6pt]\\[2pt]
}
\date{}

\setlist{nolistsep}
\begin{document}
\maketitle
\setlength{\oddsidemargin}{ 1cm}
\setlength{\evensidemargin}{\oddsidemargin}
\setlength{\textwidth}{15.50cm}
\vspace{-.8cm}
\begin{center}
\parbox{\textwidth}{
\textbf{Abstract:}\quad {
  Oscillation can be observed in candle flames under certain circumstances. This research theoretically proposed an hydrodynamic-instability-based approach towards the oscillation and coupling of candle flames, claiming that the visible flame is part of a jet flow undergoing three stages: laminar, wave, and turbulence. In the ordinary stable combustion the visible flame is in the laminar stage, while under these oscillating circumstances it is in the surface wave stage. In-phase coupling was explained as two jet flows merging into one, and anti-phase coupling occur when mergence becomes unavailable. The Schlieren image experiments, frequency measuring experiments and some ``disturbance'' experiments were conducted to support this view. Mathematical models were also proposed for description.
} \\

}
\end{center}

\setcounter{page}{1}

\setlength{\oddsidemargin}{-.5cm}  
\setlength{\evensidemargin}{\oddsidemargin}
\setlength{\textwidth}{17.00cm}
\section*{Originality}

\begin{itemize}
  \item \textbf{Creatively proposed the hydrodynamic-instability based approach} towards understanding both the cause of the oscillation of candle flame oscillators and the mechanism of their coupling;
  \item Duplicated the existing experiments while \textbf{designing and conducting original ones} to support our proposal.
\end{itemize}

\section*{Introduction}

\subsection*{Research Background}

When several candles burn next to each other, an oscillation of the flames could be observed. Further, when two or more such oscillators are placed in a proper manner, they could couple with each other, illustrating either in-, anti-phase or other forms of synchronization.

There have been debates on determining the cause of the oscillation and the mechanism by which they could couple with each other. There are generally two models concerning the cause of the oscillation: Y. Nagamine et al.\upcite{descend} proposed that when several candles burn next to each other and form an oscillator, they burn drastically. The candles are then heated to provide redundant paraffin, resulting in the accumulation of products of incomplete combustion above the flames. When these product are cooled, they form a high-density layer which, after reaching certain critical condition, fall through the flames. This model was aimed to explain the occurrence of turbulence right above the oscillating candle flames and the observed detachment of the flames during oscillation.

Another widely accepted view is that the oscillation is caused by the vortex formed next to the visible flames. In their researches, Okamoto K. et al.\upcite{vortex1} and Yang, Tao\upcite{vortex2} proposed that, as candles burn more drastically near each other, the air adjacent to the flames would ascend at a high speed. After reaching an critical speed, vortices would form where the gradient of flow velocity are relativily high and repel the flame surface. This repelling force would then stretch the flames into halves, and the lower half would continue to participate in the next oscillation. 

There are also two main theories concerning the mechanism of coupling: Take one simplest case as example. When two oscillators are put together, while close enough, they fall into an in-phase synchronization. If we add to the distance in between, for a certain range of distance the oscillators fall into an anti-phase synchronization. Further apart, their oscillation then become independent of each other. Cheng, Ting et al.\upcite{radiation} proposed in their research that, the interaction between two oscillators are mediated by thermal radiation. Suppose the influence of thermal radiation is restricted to certain region surrounding the candle flame. While the oscillators are close enough to each other so that however they oscillate, they always fall into this ``effective region'' of each other's thermal radiation, to reach the minimum potential, they would oscillate in an in-phase manner. However, when we prolong the distance in between, it can not be guaranteed that their interaction constantly exists. If one oscillator is at its minimum size, the other would have to be near its maximum size to keep their interaction. That would be the reason for anti-phase synchronization. When put further apart and no synchronization mode could keep their interaction, the system goes into chaos and the oscillators would be independent of each other.

Another theory contains a different basic assumption and claims that the coupling of oscillators is enabled by the interaction of vortices, which is an echo to the earlier ``vortex theory'' for the cause of oscillation. Okamoto, K. et al.\upcite{vortex1} proposed that, as the formed vortices have a scale of size similar to the diameter of a flame, two or more oscillators could interact with each other through the formation of vortices (for example, by repelling each other's flame surfaces), and therefore their relative position could determine the coupling mode.

In the above mentioned researches and others\upcite{kitahata,arrays,ampDeath,coFlow,collective}, experiments or simulations were also conducted to support each view.

\subsection*{Summary of Our Work}

While going through the existing researches on this topic, we found that each of the proposition focused on a local aspect and ascribed the phenomena to a single quantity, whether it's the product of incomplete combustion, local vortices, or thermal radiation. Though they suffice respectively in explaining certain phenomena, they do not form a systematic understanding of the topic.

After theoretical and experimental investigation, we propose our ``hydrodynamic-instability based approach'' which provides a wholistic understanding of the physics of this topic. We claim that the oscillation of candle flames is a case of Kelvin-Helmholtz instability: The jet flow above any single candle go from laminar flow to turbulence at some distance, while placing several candles together has prolonged the transition stage, allowing the visible part of the flame to be covered and therefore causing visible oscillation; as the properties of the instability are determined by quantities including the temperature, density and so on, if we place two or more oscillators together, by influencing these quantities they could couple with each other. In consequence, we discovered the similarity between in-phase coupled candle oscillators and a single oscillator, and proposed that we regard them as the same theoretical entity. We also applied the \textit{kuramoto model} of synchronized systems to study the transition of coupling stages.

Our theory therefore suggested the following:
\begin{enumerate}
  \item that the impact on frequency of making oscillators in-phase coupled is similar to that of increasing the number of candles in a single oscillator; [Section \ref{theo:inPhase}, \ref{exper:frequency1}, \ref{exper:frequency2}]
  \item that making oscillators anti-phase coupled would do the opposite on frequency; [Section \ref{theo:antiPhase}, \ref{exper:frequency1}, \ref{exper:frequency2}]
  \item that ``bifurcation'' occurs in transition between in- and anti-phase coupling along with a bottle-neck stage, while between anti- and non-coupling it does not; [Section \ref{theo:transition}, \ref{exper:frequency2}]
  \item that transition from in- to anti- phase coupling takes place when the distance between oscillators start to allow each to reach a full amplitude; [Section \ref{theo:antiPhase}, \ref{exper:Schlieren}]
  \item that certain local disturbance above the flame would not significantly affect the oscillation of the flame; [Section \ref{theo:disturbance}, \ref{exper:disturb}]
  \item and that certain disturbance on the flame's side could generate oscillation in an non-oscillating single candle. [Section \ref{theo:disturbance}, \ref{exper:disturb}]
\end{enumerate}

By conducting Schlieren image experiments, frequency-distance measuring experiments and disturbance experiments, we verified the above suggestions to some satisfactory extent.

\section{Theory}

\subsection{The three stages}

Consider the burning process of a candle: First, the paraffin evaporate as being heated; second, the paraffin get mixed with adjacent air; third, chemical reaction takes place, consuming the paraffin along with oxygen in the air, and produce water vapour, carbon dioxide and certain by-products. This physical system bears many similarities to a jet flow in air, with only the exact quantities, along with their changes, in flow velocity, density etc. wanting further discussion.

\begin{figure}[H]
  \centering
  \includegraphics[width=0.6\textwidth]{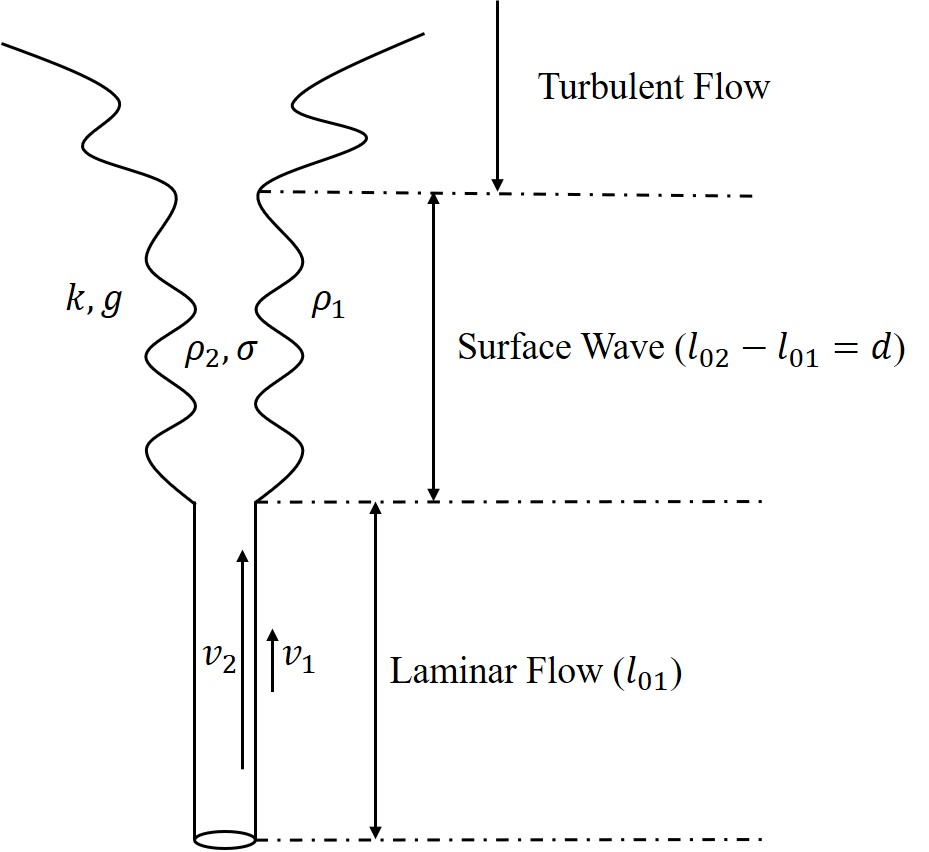}
  \caption{Illustration of the three stages.}
  \label{fig:threeStages}
\end{figure}

As shown in Fig.\ref{fig:threeStages}, the jet flow above a candle flame normally exhibit three stages of evolution: Laminar flow, surface wave and turbulence. We denote the wave number of the surface wave as $k$, the gravitational acceleration $g$, the flow velocity, density and surface tension of the jet flow $v_2, \rho_2, \sigma$, and those of the adjacent air $v_1, \rho_1$. We suppose the length of the laminar flow is $l_{01}$, the wave surface $d$, and denote the distance of certain point to the nozzle as $l$.

By hydrodynamics we propose that this surface wave is caused by Kelvin-Helmholtz instability. We may first study the local case, where we ignore the impact of $g$ and regard the other quantities as constant. We can see that, surface waves occur as the driving force of the jet flow itself and the dragging force of surface tension provided by adjacent air come to reach a critical region.

\begin{figure}[H]
  \centering
  \subfigure[The theoretical shape.]{
    \label{fig:withoutg}
    \includegraphics[width=0.4\textwidth]{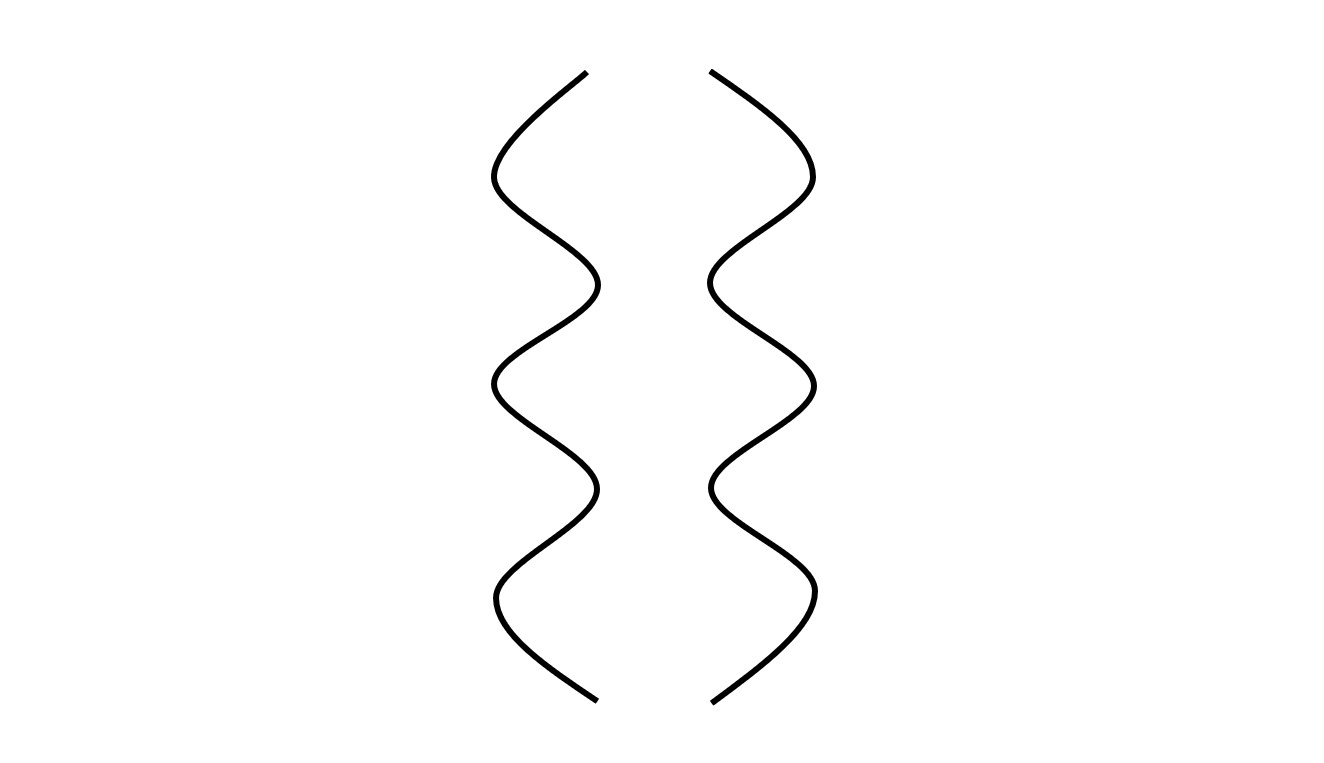}}
  \subfigure[The adjusted shape.]{
    \label{fig:withg}
    \includegraphics[width=0.4\textwidth]{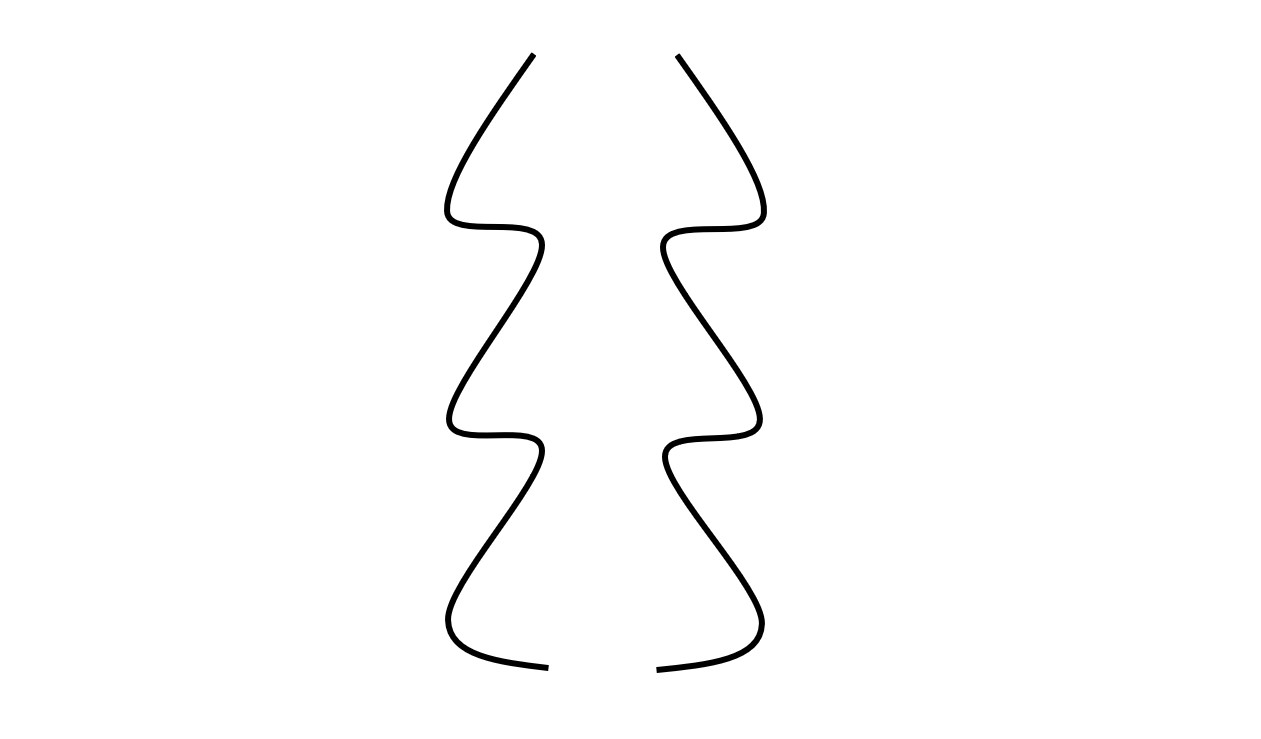}}
  \caption{Shape of surface waves.}
  \label{fig:withAndWithoutG}
\end{figure}

Taking gravitation into consideration, we then examine the assymmetry of the surface waves' shape and the jet flow's transition from laminar flow to turbulence. On the one hand, as the paraffin vapour has a larger density than air, the wave crests would have lower positions than their theoretical ones, and the wave valleys higher ones, as shown in Fig.\ref{fig:withAndWithoutG}, therefore exhibiting assymmetry. On the other hand, if we regard the surface wave stage as a transition stage between laminar flow and turbulence (and indeed it is), it's obvious that the transition is brought about by the speed-down of the jet flow with hight as a common result of gravitation and surface tension.

\subsection{Oscillation}
\label{theo:oscillation}

With the thorough discussion above, we shall first examine the conditions under which a visible oscillation takes place. Ordinarily we observe candles oscillating when
\begin{enumerate}
  \item several candles are bound together;
  \item a single candle with average-size wick has a rather large diameter;
  \item \dots
\end{enumerate}

Based on previous researches already cited in the \textbf{Introduction}, it's likely that the common trait of these conditions would be that the combustion is incomplete. This causes the accumulation of paraffin vapour and high-density by-products above the flame accompanied by the heating of a larger space of air, leading to a decrease in the general relative jet flow velocity, which has the exact consequence of prolonging the surface wave stage $d$ and shortening the laminar flow stage $l_{01}$. 

The fact that a typical candle doesn't illustrate oscillation is simply because its visible flame is in the laminar flow stage. By bouding candles together (i.e., cutting air supply) or providing superfluous paraffin, the jet flow get into the surface wave stage right above the ``nozzle'' and then the waves become visible as the flames are in their area.

\subsection{Coupling}

We shall now look into the two most frequently observed coupling modes: the in-phase coupling and the anti-phase one. We could also do some primary research on how transitions are mediated.

\subsubsection{In-phase coupling}
\label{theo:inPhase}

\begin{figure}[H]
  \centering
  \subfigure[In-phase.]{
    \label{fig:inPhase}
    \includegraphics[width=0.4\textwidth]{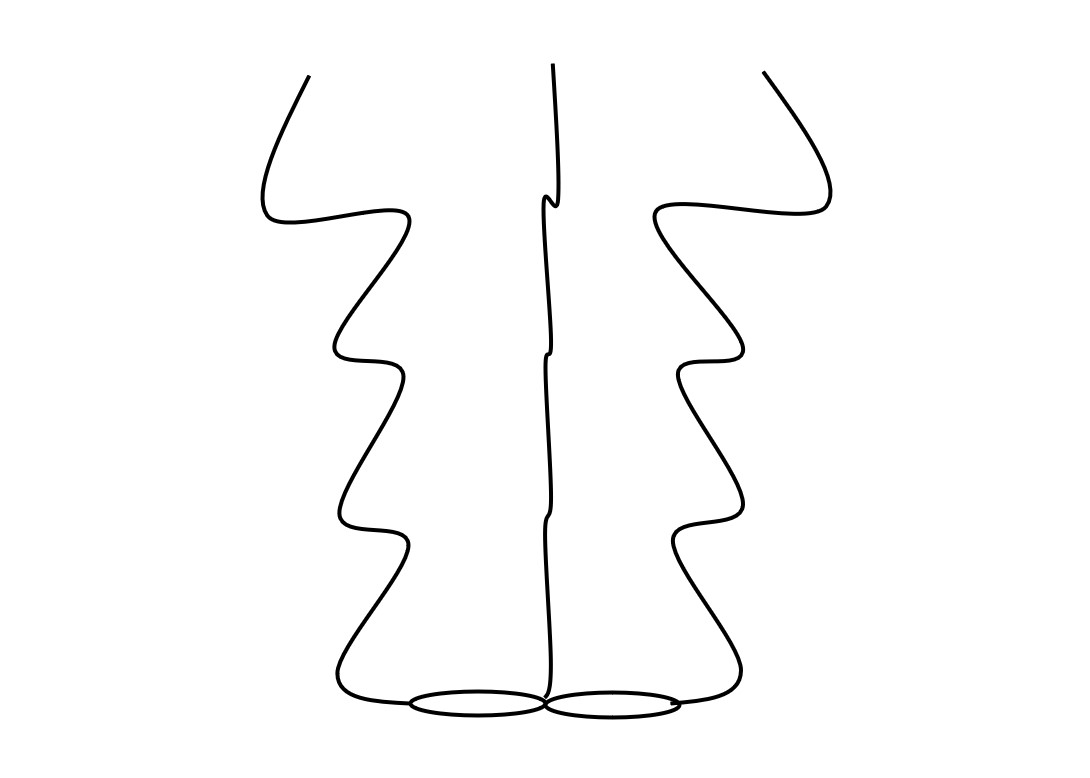}}
  \subfigure[Anti-phase.]{
    \label{fig:antiPhase}
    \includegraphics[width=0.4\textwidth]{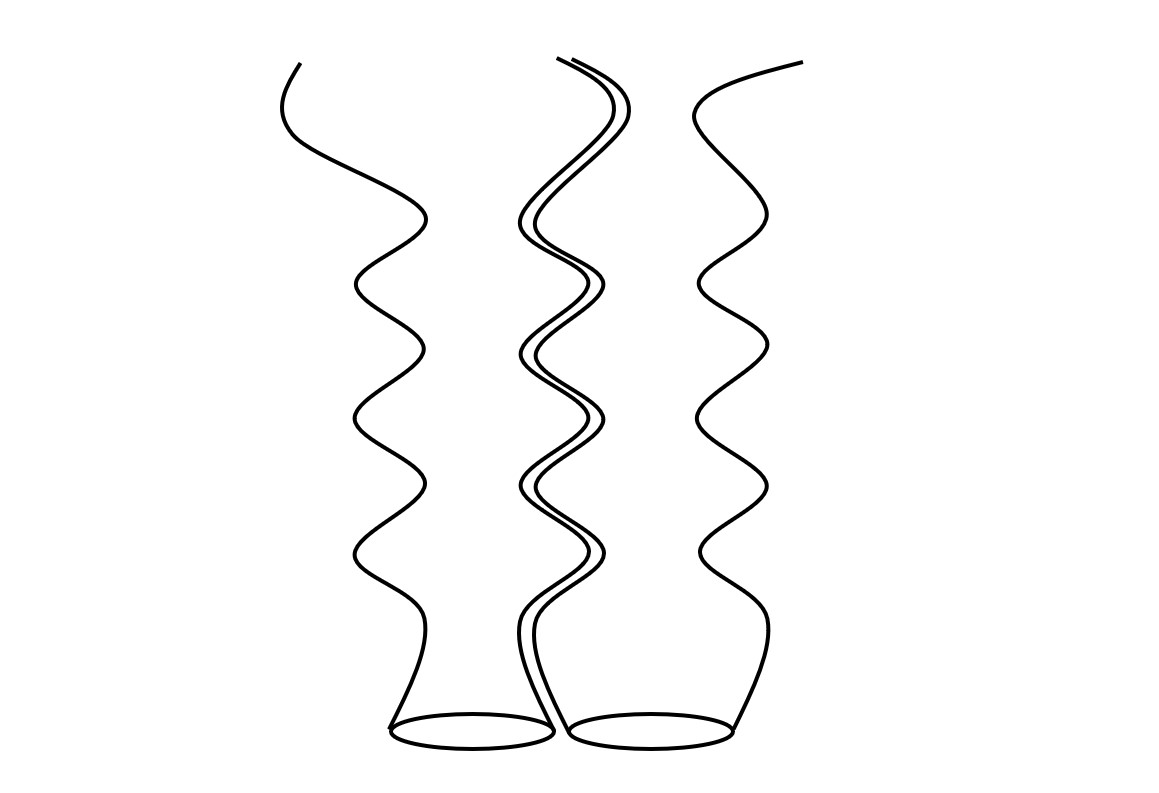}}
  \caption{In- and anti- phase coupling}
  \label{fig:inAndAntiPhase}
\end{figure}

In our theory, there is no in-phase coupling. We regard the so-called ``in-phase coupling'' as the same theoretical entity as a single oscillator. Their jet flow merge together and they share the same frequency. The physical picture could be easily understood with reference to Fig.\ref{fig:inPhase}. For experiment, their being identical could be manifested by the fact that adding candles to a single oscillator would change its frequency just the same as in-phase coupling it with another oscillator would.

\subsubsection{Anti-phase coupling}
\label{theo:antiPhase}

The theory gets tricky for anti-phase coupling. We deem that transition from in-phase coupling to anti-phase coupling occurs when the distance between two oscillators come to a critical value where each oscillator is able to reach its full amplitude. At this stage to merge and form a single jet flow would requires higher order and the system naturally falls into anti-phase synchronization, where the ``touching'' surface of the two jet flows stay parallel to each other. In this case, the flow velocity difference of the jet flow and adjacent air (local but not general environment) in the middle would be rather small, and each single oscillator would have a higher flow velocity compared to when it oscillates alone.

As illustrated in Ref.\cite{tempr}, the temperature distribution of air adjacent to a candle flame is practically similar to a Gauss distribution with its standard deviation being $1.23$ times the radius of the jet flow:
\begin{equation}
  \frac{T - T_0}{T_A - T_0} = e^{- \frac{r^2}{\sigma^2}},
\label{eq:temperature1}
\end{equation}
with $T_0$ as the background temperature, $T_A$ the maximum temperature of the flame, $r$ the horizontal distance to the flame center and $\sigma$ the standard deviation. If we then express the interaction between two jet flows as a superposition of two such distributions, we have:
\begin{equation}
  \frac{T - T_0}{T_A - T_0} = e^{- \frac{(x - d)^2}{\sigma^2}} + e^{- \frac{(x + d)^2}{\sigma^2}},
\label{temperature2}
\end{equation}
where $2d$ is the distance between two oscillators. When $d < 0.7\sigma$, the combined distribution has only one peak, while when $d > 0.7\sigma$ it begins to have two peaks. This critical value of $d$ is proposed to be the transition distance of in- and anti- phase coupling.

By Sec.\ref{theo:oscillation} and \ref{theo:inPhase}, concerning the impact on the frequency we have [bounding several candles together] $\sim$ [increasing the diameter of a single candle] $\sim$ [in-phase coupling oscillators], and we already know the first two would decrease the relative flow velocity. We therefore propose that by anti-phase coupling oscillators (i.e., incresing relative flow velocity against the general environment) the impact on frequency would be quite the opposite.

\subsubsection{Transition}
\label{theo:transition}

In Ref.\cite{arrays}, synchronizing systems of multiple oscillators have been examined using the \textit{kuramoto model}. Though the model is intended for the behavior of a large set of coupled oscillators, we find it efficient in discussing transition of stages in our own theory.

Assume a system with two identical oscillators with intrinsic frequency $\omega$, we have
\begin{eqnarray}
  && \dot{\theta_1} = \omega + K \sin(\theta_2 - \theta_1),\\
  && \dot{\theta_2} = \omega + K \sin(\theta_1 - \theta_2),
\end{eqnarray}
with $K$ as the coupling strength and $\theta_i$ the phase of the $i^{th}$ oscillator. The above equations can also be rewritten as
\begin{eqnarray}
  && \dot{\theta_1} + \dot{\theta_2} = 2 \omega,\\
  && \dot{\theta_1} - \dot{\theta_2} = - 2 K \sin(\theta_1 - \theta_2).
\label{eq:dotAlpha}
\end{eqnarray}

We could define phase difference $\alpha = \theta_1 - \theta_2$ to get
\begin{eqnarray}
  && \theta_1 = \omega t + \frac{\varphi}{2} + \frac{1}{2}\alpha,\quad \theta_2 = \omega t + \frac{\varphi}{2} - \frac{1}{2}\alpha,\\
  && \alpha = \arccos \frac{(1 + \cos \alpha_0)e^{4Kt} - (1 - \cos \alpha_0)}{(1 + \cos \alpha_0)e^{4Kt} + (1 - \cos \alpha_0)}.
\end{eqnarray}

As $t \to +\infty$, when $K>0$, $\alpha \to 0$, the oscillators are in-phase coupled; when $K<0$, $\alpha \to \pi$, the oscillators are anti-phase coupled; and when $K=0$, $\alpha = \alpha_0$, the oscillators are de-coupled. Therefore in our case, $K$ must be a function of the distance between the oscillators which determines how the oscillators couple with each other.

We graph $\dot{\alpha}$ to $\alpha$ to find the stable fixed points on the $\alpha$ axis. As Eq.\ref{eq:dotAlpha} shows, $\dot{\alpha} = - 2 K \sin \alpha$:

\begin{figure}[H]
  \centering
  \subfigure[$K>0$.]{
    \label{fig:dotAlpha1}
    \includegraphics[width=0.4\textwidth]{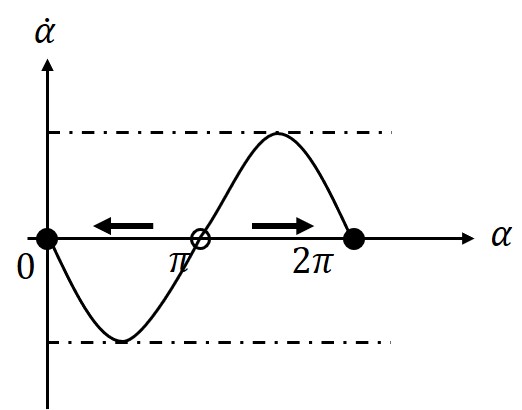}}
  \subfigure[$K<0$.]{
    \label{fig:dorAlpha2}
    \includegraphics[width=0.4\textwidth]{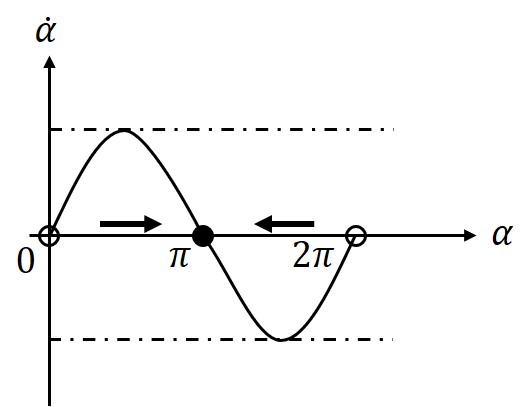}}
  \caption{Fixed points for phase difference $\alpha$.}
  \label{fig:dotAlpha1and2}
\end{figure}

As Fig.\ref{fig:dotAlpha1and2} show, when $K$ transitions from positive to negative, the dotted lines approach the $\alpha$ axis and cause the stable fixed point of the axis change from $0$ suddenly to $\pi$ ($\alpha \in [0, 2 \pi)$). This is called a bifurcation. 

We could also notice that, during the transition, as $K \to 0$, $\dot{\alpha}$ also $\to 0$, therefore the process of re-coupling is a rather slow one. As our model has no statement about how the transition from a single jet flow to two takes place, there may be abnormal stages during the process.

In the ultimate stage where the two oscillators are totally seperated, $K \to 0$, suggesting no bifurcation during transition between anti-phase coupling and de-coupling because the sign of $K$ doesn't change.

\subsection{Disturbance}
\label{theo:disturbance}

We also designed additional experiments to help verify our theory, and we shall explain the purpose here in the \textbf{Theory} section. We call these experiments the ``disturbance'' experiments.

From our three-stage theory, as both the driving and dragging force are either from below or from the sides, it naturally follows that, if we change the condition above, it would not significantly change the phenomena. Oscillation would still occur with the same frequency. On the contrary, in a system where no oscillation is visible, say, a single average candle, we could produce oscillation by increasing its relative jet flow velocity without adding candles, therefore preventing the interference of heat.

\section{Experiment}

To illustrate the validity of our theory, we conducted the following experiments.

\subsection{Schlieren Image Experiments}
\label{exper:Schlieren}

\begin{figure}[H]
  \centering
  \includegraphics[width=0.8\linewidth]{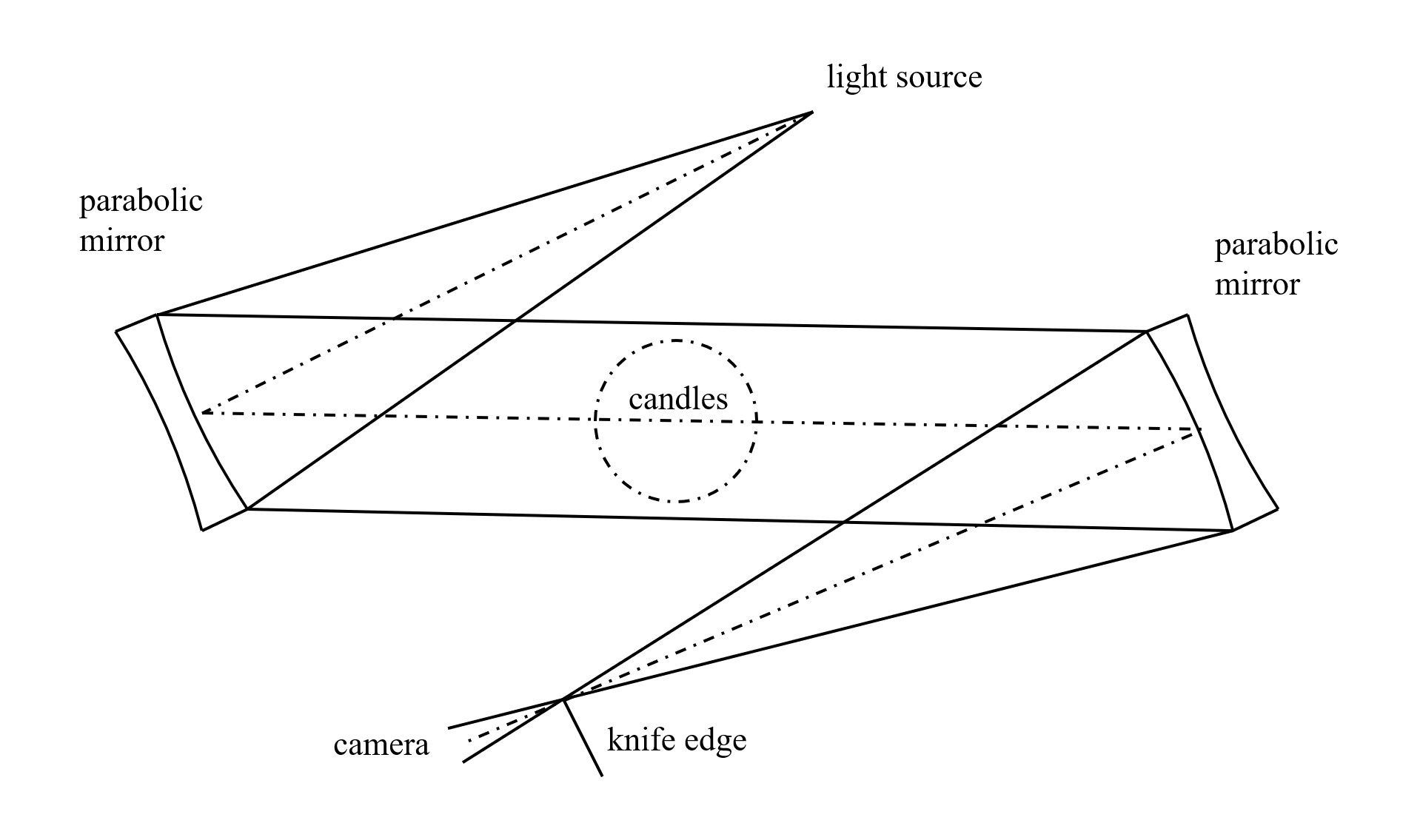}
  \caption{Schlieren Image setup.}
  \label{fig:schlierenSetup}
\end{figure}

\textit{Schlieren imaging} is a common way to visualize air flows.\upcite{radiation} Our experimental setup is as shown in Fig.\ref{fig:schlierenSetup}

\begin{figure}[H]
  \centering
  \subfigure[A single oscillator.]{
    \label{fig:schlierenSingle}
    \includegraphics[width=0.3\textwidth]{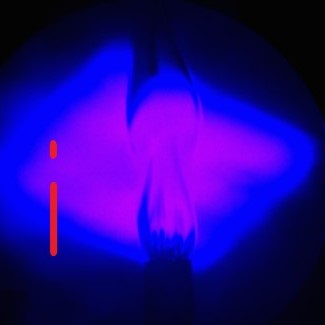}}
  \subfigure[In-phase coupled oscillators.]{
    \label{fig:schlierenIn}
    \includegraphics[width=0.3\textwidth]{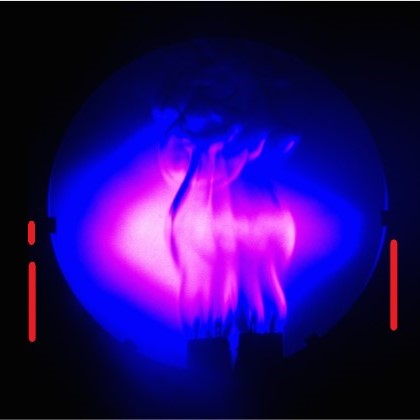}}
  \subfigure[Anti-phase coupled oscillators.]{
    \label{fig:schlierenAnti}
    \includegraphics[width=0.3\textwidth]{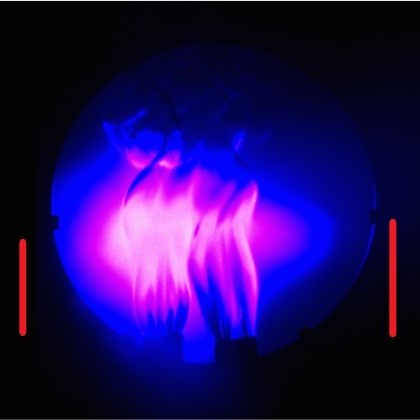}}
  \caption{Schlieren images of oscillators.}
  \label{fig:schlierenOscillators}
\end{figure}

And the typical schlieren images of a single oscillator, of in-phase coupled and of anti-phase coupled oscillators are as shown in Fig.\ref{fig:schlierenOscillators}. The shape of surface wave in Fig.\ref{fig:schlierenSingle} is just as illustrated in Fig.\ref{fig:withg}, and Fig.\ref{fig:schlierenIn}, Fig.\ref{fig:schlierenAnti} just as Fig.\ref{fig:inPhase}, Fig.\ref{fig:antiPhase}. The red bars denote the range of visible flames.

For each of the three cases, we set the parameters constant and measure the maximum width of the jet flow using the total width of the bound candles (an average of $23 mm$) as a reference. Note that we chose the anti-phase coupling case to have the minimum distance between oscillators. We then have Table \ref{table:schlieren}, where $\sigma_i$ is the standard deviation. This verifies our proposition that the coupling mode turn from in-phase to anti-phase when both of the oscillators could reach their full amplitudes.

\begin{table}[H]
  \centering
  \caption{Scales of jet flows}
  \label{table:schlieren}
  \begin{tabular}{c|c|c|c|c|c|c}
    \hline
    \hline
    \quad & Single & $\sigma_1$ & In-Phase & $\sigma_2$ & Anti-Phase & $\sigma_3$ \\ \hline
    Maximum Width/mm & $38.6$ & $0.8$ & $64.2$ & $2.31$ & $37.6$ each & $1.35$ \\
    \hline 
    \hline
  \end{tabular}
\end{table}

Also, by our ``Gauss-distribution'' theory, we can calculate the critical $2d$ as ($D$ is the diameter of an oscillator):
\begin{equation}
  2d = 2 \times 0.7 \times 1.23 D = 2 \times 0.7 \times 1.23 \times (\frac{38.6}{2}) \approx 33\text{mm},
\end{equation}
which is exactly the observed mean distance of transitions.

\subsection{Frequency-Distance Measuring Experiments}

Frequency is another measurable quantity that reveals the intrinsic property of an oscillating system.

\subsubsection{For a single oscillator}
\label{exper:frequency1}

We first change the number of candles in a single oscillator and investigate the change in its intrinsic frequency. We also tried two types of configurations: to put them in a straight line or to put them in a most compact manner. For there is a detailed discussion in Ref.\cite{radiation}, we only did basic experiments to check the results.

\begin{figure}[H]
  \centering
  \subfigure[2 candles.]{
    \label{fig:2}
    \includegraphics[width=0.18\textwidth]{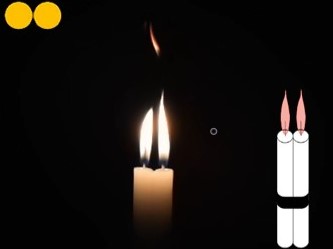}}
  \subfigure[3 candles (linear).]{
    \label{fig:3line}
    \includegraphics[width=0.18\textwidth]{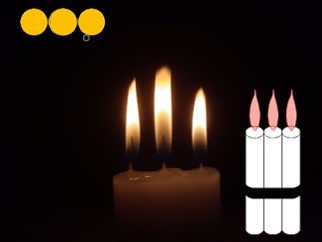}}
  \subfigure[3 candles (compact).]{
    \label{fig:3compact}
    \includegraphics[width=0.18\textwidth]{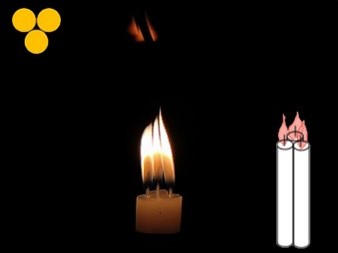}}
  \subfigure[4 candles (linear).]{
    \label{fig:4line}
    \includegraphics[width=0.18\textwidth]{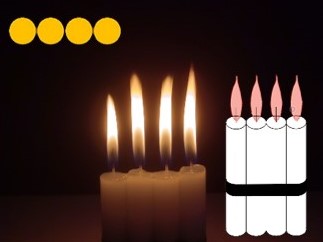}}
  \subfigure[4 candles (compact).]{
    \label{fig:4compact}
    \includegraphics[width=0.18\textwidth]{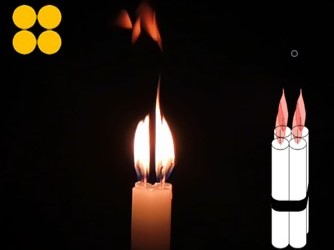}}
  \caption{Oscillators with different numbers of candles.}
  \label{fig:number}
\end{figure}

\begin{figure}[H]
  \centering
  \subfigure[2 candles.]{
    \label{fig:schlieren2}
    \includegraphics[width=0.18\textwidth]{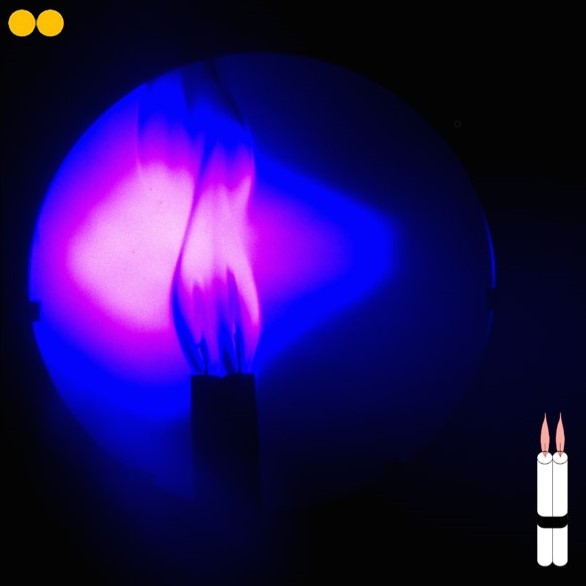}}
  \subfigure[3 candles (linear).]{
    \label{fig:schlieren3line}
    \includegraphics[width=0.18\textwidth]{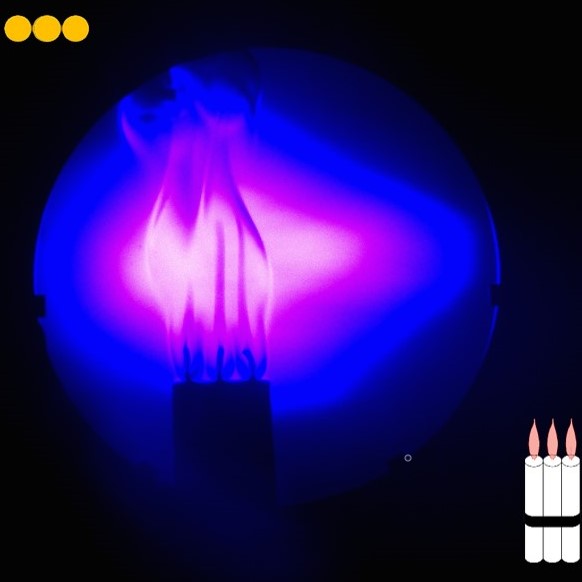}}
  \subfigure[3 candles (compact).]{
    \label{fig:schlieren3compact}
    \includegraphics[width=0.18\textwidth]{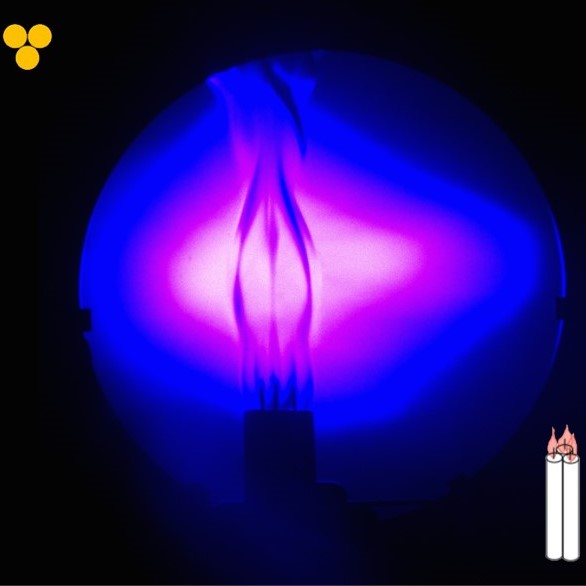}}
  \subfigure[4 candles (linear).]{
    \label{fig:schlieren4line}
    \includegraphics[width=0.18\textwidth]{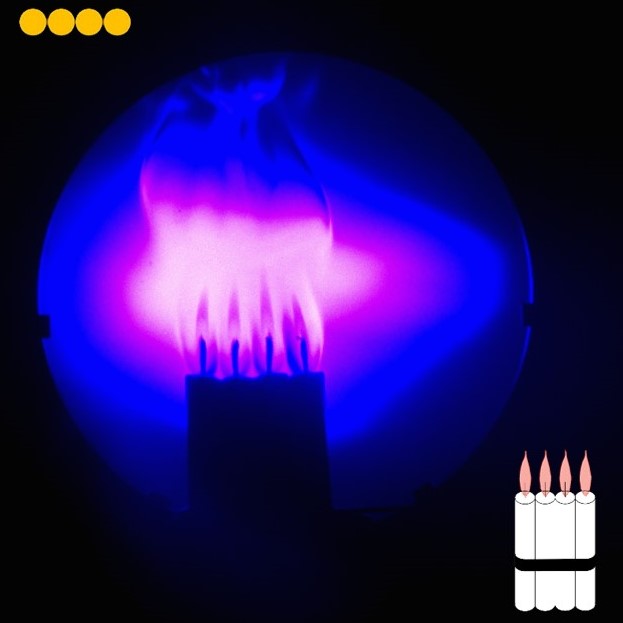}}
  \subfigure[4 candles (compact).]{
    \label{fig:schlieren4compact}
    \includegraphics[width=0.18\textwidth]{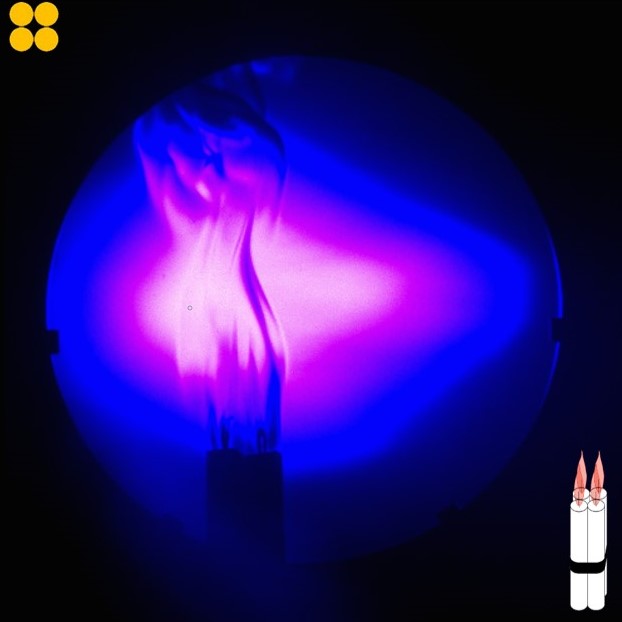}}
  \caption{Schlieren images of oscillators with different numbers of candles.}
  \label{fig:schlierenNumber}
\end{figure}

\begin{figure}[H]
  \centering
  \subfigure[The linear configuration.]{
    \label{fig:freqLinear}
    \includegraphics[width=0.45\textwidth]{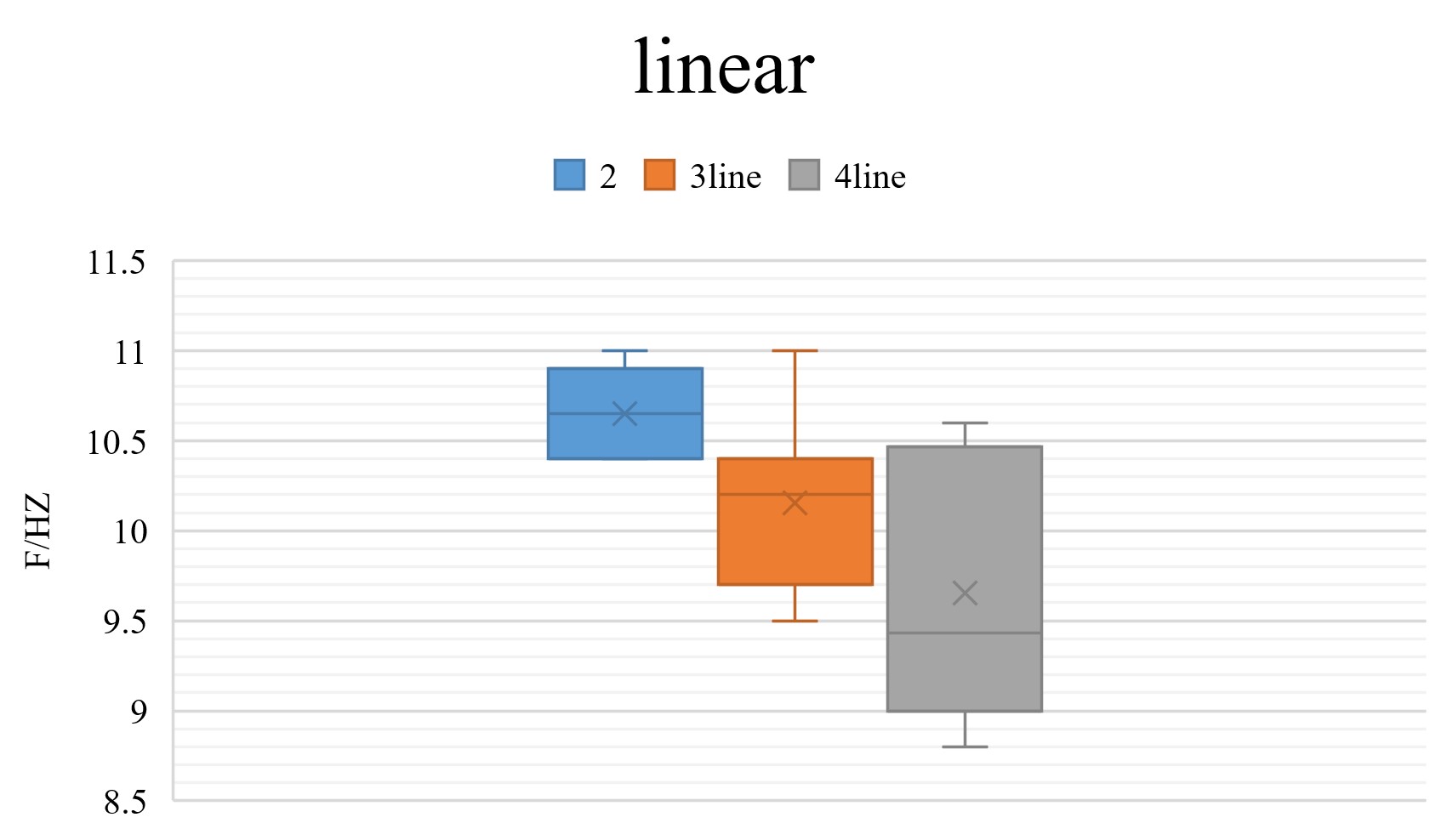}}
  \subfigure[The compact configuration.]{
    \label{fig:freqCompact}
    \includegraphics[width=0.45\textwidth]{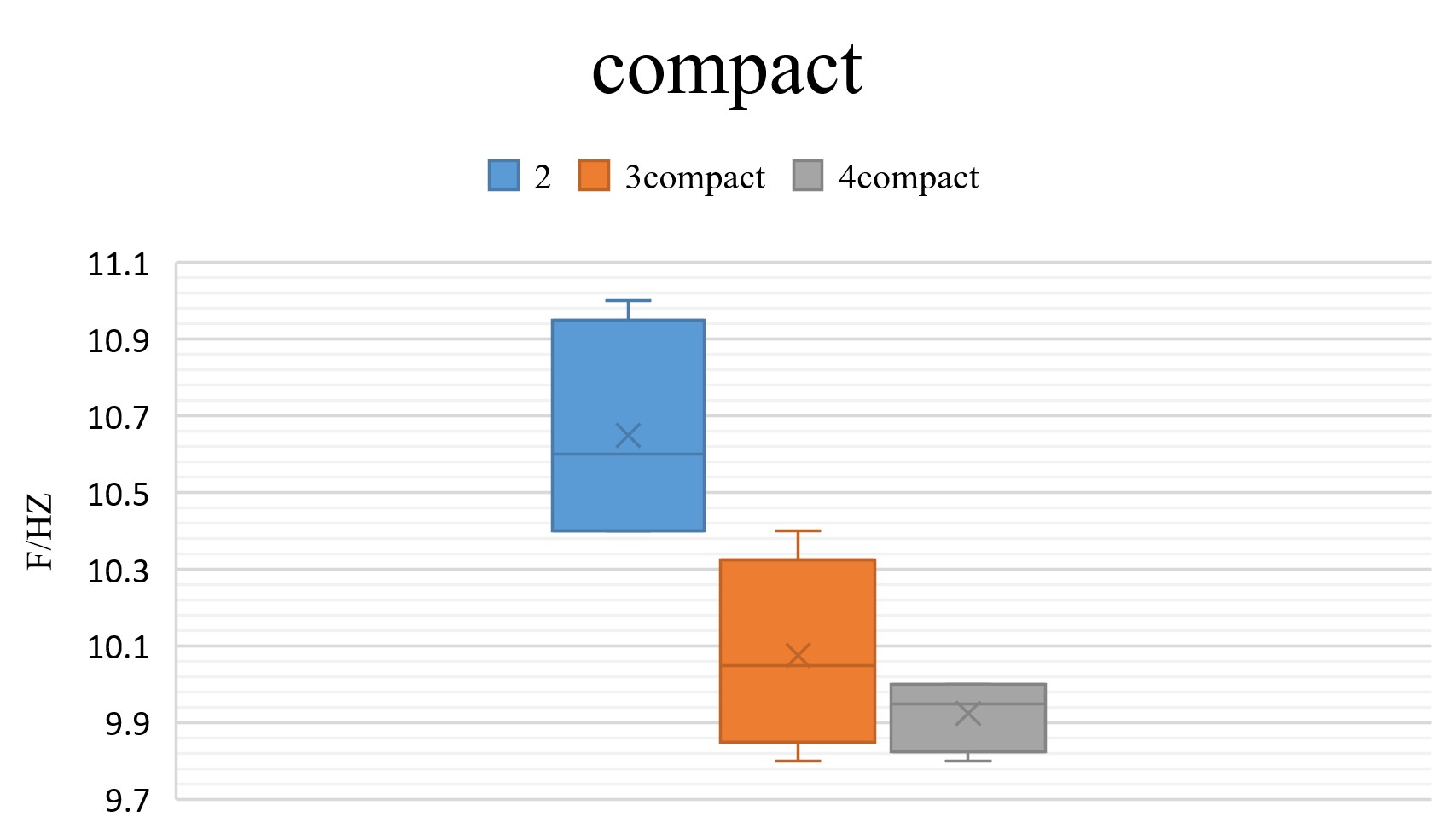}}
  \caption{Frequency-Number of Candles relation for a single oscillator.}
  \label{fig:freqSingle}
\end{figure}

From the experimental results above we could conclude the following:
\begin{enumerate}
  \item In the range of number of candles considered, the more candles in an oscillator the lower the frequency.
  \item A compact configuration would have relatively higher frequency than a linear configuration of the same number of candles.
\end{enumerate}

From Sec.\ref{theo:antiPhase} we then shall examine whether an in-phase coupling could lead to a decrease in frequency and an anti-phase coupling an increase. The next experiment serves this purpose well.

\subsubsection{For coupled oscillators}
\label{exper:frequency2}

\begin{figure}[H]
  \centering
  \subfigure[The setup.]{
    \label{fig:distanceSetup}
    \includegraphics[width=0.35\textwidth]{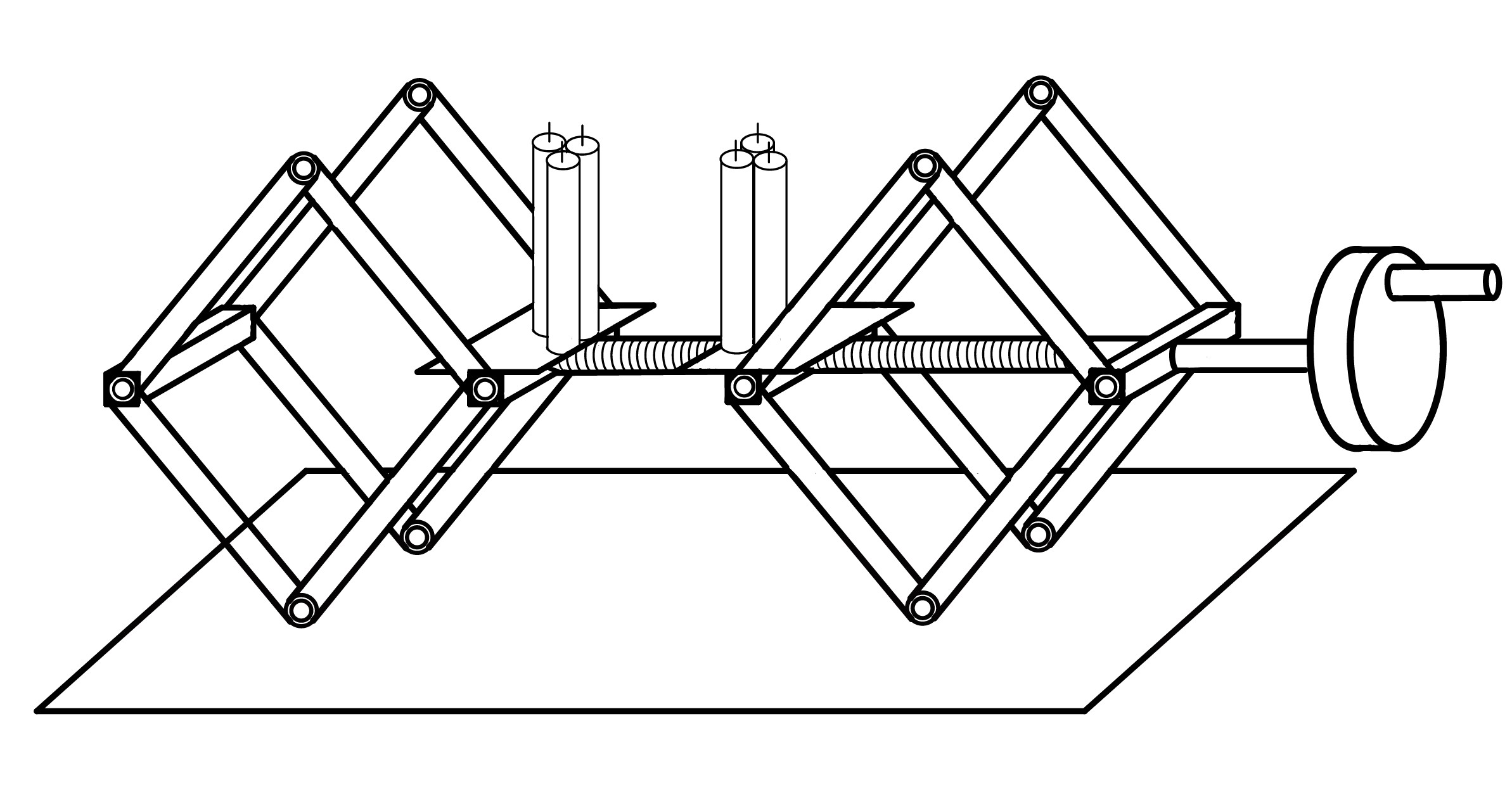}}
  \subfigure[Frequency-Distance relation for a two-oscillator system.]{
    \label{fig:freqDistance}
    \includegraphics[width=0.55\textwidth]{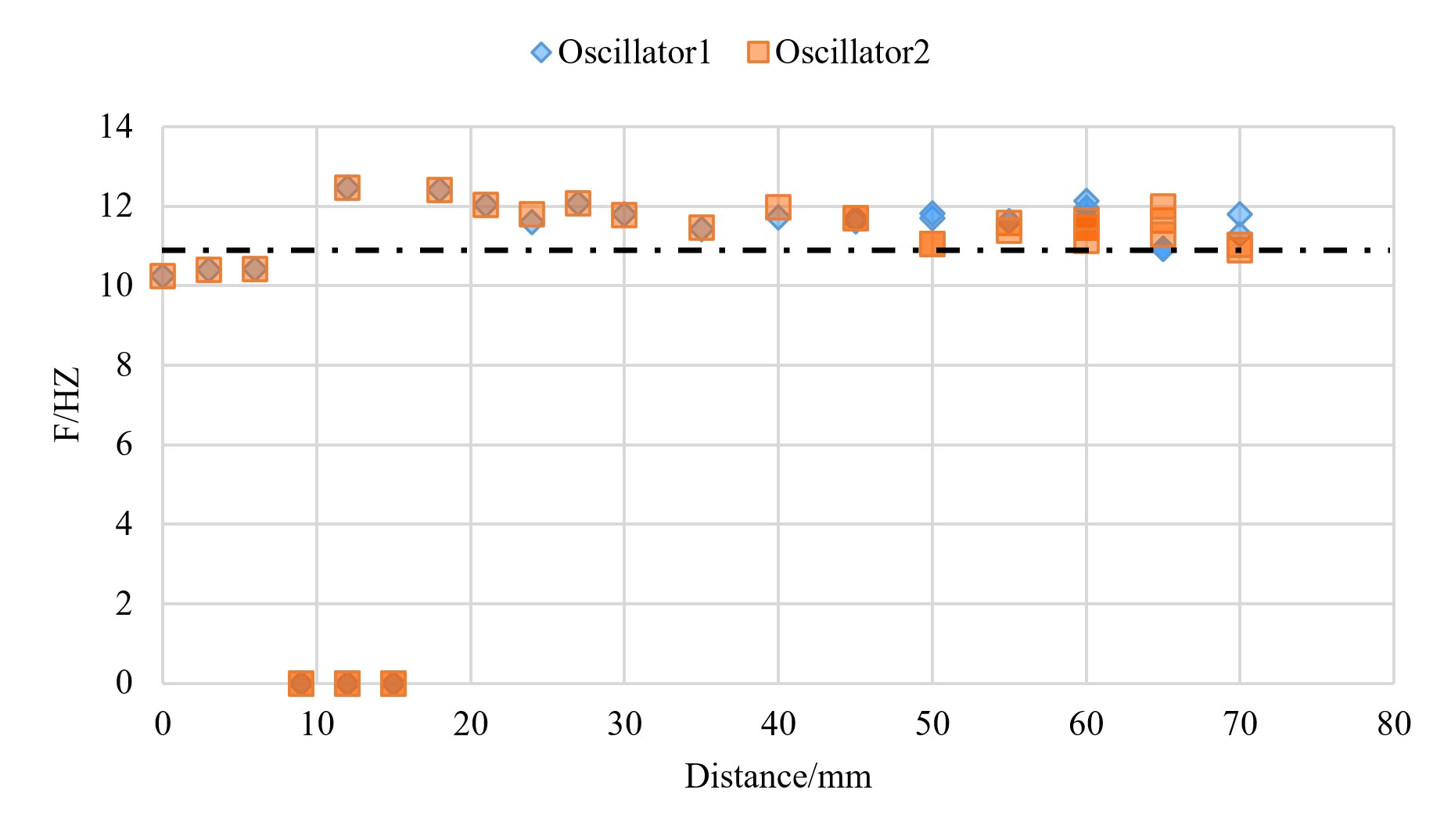}}
  \caption{The Frequency-Distance experiment.}
  \label{fig:Distance}
\end{figure}

We then measure the frequency-distance relation of a two-oscillator system of oscillators consisting of three candles to check the well-known results. The experimental setup and the final results are as follow (Fig.\ref{fig:freqDistance}). Frequency was measured by: converting each frame of the experiment video (filmed with a high-speed camera of 480 fps) into its gray-scale version, settinng a critical gray value, calculating the number of valid pixels as a function of time and using DFT (discrete Fourier transform) to obtain the final result.

Again, from the experimental results above we could conclude the following:
\begin{enumerate}
  \item The in-phase coupling indeed causes a decrease in the frequency.
  \item The anti-phase coupling, on the other hand, causes a increase in the frequency.
  \item The transition between in-phase and anti-phase coupling is a sudden one.
  \item The transition between anti-phase coupling and non-coupling is a gradual one.
  \item An abnormal stage do occur during transition from in- to anti-coupling, which is typically called ``the Amplitude Death''.\upcite{vortex1,vortex2,radiation,ampDeath,coFlow}
\end{enumerate}

\subsection{Disturbing the Oscillators}
\label{exper:disturb}

As explained in Sec.\ref{theo:disturbance}, we shall conduct two disturbance experiments: one from above, and one from the sides. We first use a rectangle-shaped piece of glass to ``cut off'' the upper part of the flame of an oscillator containing three candles, and obtain the disturbed oscillator's frequency. Then we place the same glass on a single candle's side at different distances to observe whether it would oscillate. Before each experiment the glass was cooled to an equilibrium temperature. 

\begin{figure}[H]
  \centering
  \subfigure[An oscillator disturbed from above.]{
    \label{fig:above}
    \includegraphics[width=0.20\textwidth]{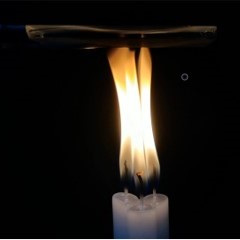}}
  \subfigure[Schlieren image of an oscillator disturbed from above.]{
    \label{fig:schlierenAbove}
    \includegraphics[width=0.20\textwidth]{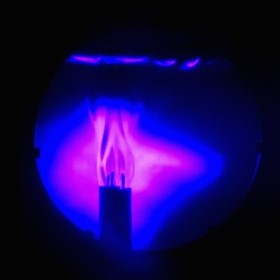}}
  \subfigure[A single candle disturbed from one side.]{
    \label{fig:side}
    \includegraphics[width=0.20\textwidth]{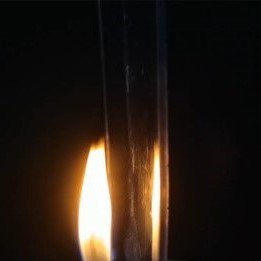}}
  \subfigure[Schlieren image of a single candle disturbed from one side.]{
    \label{fig:schlierenSide}
    \includegraphics[width=0.20\textwidth]{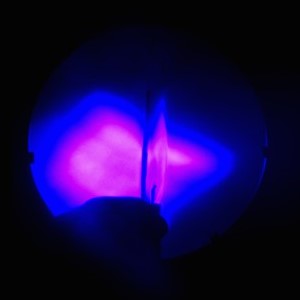}}
  \caption{The disturbance experiments.}
  \label{fig:Disturbance}
\end{figure}

Along with Fig.\ref{fig:Disturbance}, we also have:
\begin{itemize}
  \item The average frequency of oscillators disturbed from above was $10.80 Hz$, while the average frequency of undisturbed oscillators of the same kind was $10.81 Hz$.
  \item At certain distance the glass can cause a single candle to oscillate, although its oscillation is so unsteady that we have not yet been able to measure its frequency.
\end{itemize}

\section{Conclusion}

Therefore, by conducting the Schlieren image experiments, the frequency measurement experiments and the disturbance experiments, we successfully verified our theory's predictions
\begin{enumerate}
  \item that the impact on frequency of making oscillators in-phase coupled is similar to that of increasing the number of candles in a single oscillator; [Section \ref{theo:inPhase}, \ref{exper:frequency1}, \ref{exper:frequency2}]
  \item that making oscillators anti-phase coupled would do the opposite on frequency; [Section \ref{theo:antiPhase}, \ref{exper:frequency1}, \ref{exper:frequency2}]
  \item that ``bifurcation'' occurs in transition between in- and anti-phase coupling along with a bottle-neck stage, while between anti- and non-coupling it does not; [Section \ref{theo:transition}, \ref{exper:frequency2}]
  \item that transition from in- to anti- phase coupling takes place when the distance between oscillators start to allow each to reach a full amplitude; [Section \ref{theo:antiPhase}, \ref{exper:Schlieren}]
  \item that certain local disturbance above the flame would not significantly affect the oscillation of the flame; [Section \ref{theo:disturbance}, \ref{exper:disturb}]
  \item and that certain disturbance on the flame's side could generate oscillation in an non-oscillating single candle. [Section \ref{theo:disturbance}, \ref{exper:disturb}]
\end{enumerate}

And thus we have proposed a satisfactory wholistic ``hydrodynamic-instability based'' theory on the oscillation and coupling of candle flames: Consider the candle flame as the visible part of the jet flow which has three stages-- the laminar flow stage, the surface wave stage and the turbulence stage. The surface flow stage is caused by Kelvin-Helmholtz instability. A typical candle has its flame in the laminar flow stage, while an oscillator has its flame in the surface wave stage due to a slower jet flow velocity. While two oscillators are close enough for their jet flow to merge into one, they exhibit in-phase coupling. On the contrary, while the distance between them allow each of them to reach their maximum amplitude but not merge, they exhibit anti-phase coupling.

We wish the wholistic view we present here to be a seminal one because there are still more work left to be finished. For the three stages of the jet flow, inserting the height-distribution of $\rho_1, \rho_2, \sigma, v_2$ and $v_1$ into the governing equations or inequalities, one may obtain the analytical expression of the division of the three stages and certain properties of the surface wave. One could then investigate how changes in parameters might affect oscillation, and compare theoretical results with experimental data. One could then further investigate in detail how anti-phase coupling takes place. We sincerely hope that our view would be inspiring for researchers in this area.

\section{Acknowledgment}

We'd like to express our gratitude towards Nanjing University National Demonstration Center For Experimental Physics Education for providing equipment.

\section{References}
\small
\begingroup
\renewcommand{\section}[2]{}

\endgroup
\normalsize
\newpage

\end{document}